# Control of tetrahedral coordination and superconductivity in FeSe$_{0.5}$Te$_{0.5}$ thin films

S. X. Huang,[1] C. L. Chien,[1] V. Thampy,[1,2] and C. Broholm[1,2,3]

[1]*Department of Physics and Astronomy and* [2]*Institute for Quantum Matter, The Johns Hopkins University, Baltimore, Maryland 21218, USA*

[3]*NIST Center for Neutron Research, National Institute of Standards and Technology, Gaithersburg, Maryland 20899, USA*

Abstract:

We demonstrate a close relationship between superconductivity and the dimensions of the Fe-Se(Te) tetrahedron in FeSe$_{0.5}$Te$_{0.5}$. This is done by exploiting thin film epitaxy, which provides controlled biaxial stress, both compressive and tensile, to distort the tetrahedron. The Se/Te height within the tetrahedron is found to be of crucial importance to superconductivity, in agreement with the theoretical proposal that ($\pi$, $\pi$) spin fluctuations promote superconductivity in Fe superconductors.

The discovery of Fe based superconductors (Fe-SC) in 2008 [1-3] has captured intense attention from both experimentalists and theorists. To the surprise of many, the pairing symmetry of the Fe-SC appears to be not *d*-wave as in high $T_C$ cuprates, but *s*-wave [4]. One leading theoretical proposal for Fe-SCs has $s_\pm$ pairing symmetry [5-8] with two superconducting gaps of opposite sign. The mechanism for Fe based superconductivity appears to be different from previously known superconductors and rooted in their crystal structures [6, 9-11]. Fe-SCs contains a puckered FeAs(Se) sheet consisting of a two-dimensional Fe square lattice with the As (or Se) atoms located at the square centers protruding alternatively above and below the Fe plane. Unique among the Fe-SCs is the FeSe ("11") superconductor which contains only the essential FeSe sheet without interlayer as in its Fe-pnictide cousins.

The puckered FeAs(Se) sheets in all Fe-SCs consist of Fe-As(Se) tetrahedrons. The tetrahedral configuration is of crucial importance to Fe-SCs and can be advantageously investigated in FeSe. FeSe has a tetragonal crystal structure with lattice parameters *a* and *c* as shown in the inset to Fig. 1(a). The four Se atoms located above and below the Fe plane at a height of $h = zc$ form a tetrahedron characterized by the Fe-Se bond length of $(a^2/4 + z^2c^2)^{1/2}$, the Se-Fe-Se bond angle of $\alpha = 2tan^{-1}(a/2zc)$, the Se-Se interlayer separation in the *c*-direction of $c_{Se-Se} = c(1-2z)$, and the Se-Se distance of $d_{Se-Se} = [a^2/2 + c_{Se-Se}^2]^{1/2}$.

Our goal is to examine the relationship between these tetrahedral dimensions and superconductivity for a single composition. High-pressure measurements and strained epitaxial thin films are possible methods but not chemical substitution, which alters the electronic properties in addition to structure. As shown in the inset to Fig. 1(a), distortion of the tetrahedron can be achieved if the lattice constant $c$ can be decreased (increased) and $a$ be increased (decreased) from the equilibrium values. This requires biaxial stress in the basal plane, both tensile and compressive, which can be realized in epitaxial thin films but not in high-pressure measurements. In this work, we have exploited epitaxial growth of $FeSe_{0.5}Te_{0.5}$ thin films to address the crucial roles of the tetrahedral coordination. We find that the superconducting transition temperature in $FeSe_{0.5}Te_{0.5}$ thin films is directly linked to the chalcogen height $z_c$, as predicted [9] if $(\pi,\pi)$ spin fluctuations promote superconductivity.

Superconducting $FeSe_{1-x}Te_x$ thin films deposited on $SrTiO_3$, $LaAlO_3$, and $MgO$ substrates were recently been reported [12-16]. In our case, 400($\pm$50) nm thin films were grown on (100)MgO with a range of substrate temperatures ($T_S$) by pulsed laser (248 nm KrF excimer) ablation of $FeSe_{0.5}Te_{0.5}$ targets which were fabricated by two-step process[17]. The $\theta/2\theta$ x-ray diffraction pattern (Fig. 1a) shows only the (00$l$) peaks illustrating that all the $FeSe_{0.5}Te_{0.5}$ films have exclusively $c$-axis orientation. The positions of the (00$l$) peaks define the lattice parameter $c$. As shown in Fig. 1(b), with increasing $T_S$, the lattice parameter $c$ decreases monotonically. Remarkably, $c$ is *linearly* dependent on $T_S$ as $c(Å) = - T_S/T_o + d$, where $T_o = 1268$ °C and $d = 6.228$ Å. The lattice parameter $a$ can be deduced from the $d$-spacing of the (101) peak at tilted geometry via the relation of $1/d_{101}^2 = 1/a^2 + 1/c^2$. Measurements of the (200) peak and the (00$l$) peaks for two of the samples by neutron diffraction gave consistent values for $a$ and $c$ as shown in Fig. 1b. The lattice parameter $a$ increases quasi linearly with $T_S$ while maintaining the unit cell volume unchanged to within 1%. The deposition temperature $T_S$ thus avails itself as a unique tool to apply biaxial stress resulting in a linear strain of the epitaxial thin films. This is distinct from high-pressure experiments [18, 19], where both *a*, *c,* and the volume are reduced from ambient values, and from substitution experiments [10, 20-22], where the lattice constant *c* generally scales with *a*. Indeed, as shown in Fig. 1d (solid circles), epitaxial FeSe$_{0.5}$Te$_{0.5}$ thin films allow exploration of regions of lattice constants for a single composition, which cannot be accessed by other means.

The substrate temperature $T_S$ plays a key role in epitaxy, which we determined using pole-figures and in-plane $\phi$ scan about the (101) peak. Despite a significant lattice mismatch (~11%) between FeSe$_{0.5}$Te$_{0.5}$ and the MgO substrate, the films deposited at $T_S \geq 280$ °C exhibit a specific in-plane orientation with respect to the (100)MgO substrate. For $T_S \approx 500$ °C, the thin films are exclusively <100>FeSe$_{0.5}$Te$_{0.5}$||<100>MgO with excellent epitaxy (FWHM of $\Delta\phi \sim 2°$) (Fig. 2a). This epitaxy evolves into a mixture of an increasing amount of <100>FeSe$_{0.5}$Te$_{0.5}$||<110>MgO at $T_S \approx 400$ °C (Fig. 2b). At $T_S = 280$ °C, the basal plane of FeSe$_{0.5}$Te$_{0.5}$ is rotated by 45° around *c* with respect to that of MgO (<100>FeSe$_{0.5}$Te$_{0.5}$||<110>MgO) as indicated by Fig.2c. For $T_S \leq 180$ °C (Fig. 2d) only non-epitaxial growth is observed.

The conducting and superconducting properties of the FeSe$_{0.5}$Te$_{0.5}$ films also depend sensitively on $T_S$ as shown in Fig. 1(d). Samples deposited at high $T_S$ (> 400 °C) show semiconducting behavior, while those deposited at lower $T_S$ (< 400 °C) are metallic with similar resistivity traces and superconducting transition temperatures $T_C$. More importantly, the resistivity ratio and $T_C$ vary systematically but *not* monotonically with $T_S$. Starting with $T_S \approx 180$ °C, the onset of superconductivity ($T_C^{onset}$) gradually shifts to higher temperature with increasing $T_S$ as does $T_C$, the temperature at which full resistive superconductivity is achieved. The maximum values of $T_C^{onset} = 15$ K and $T_C = 11$ K are reached in the film with $T_S \approx 310$ °C. While improving epitaxy, increasing $T_S$ however degrades superconductivity and eventually produces semiconducting films.

We shall now show that the dramatic changes in the conducting properties of the FeSe$_{0.5}$Te$_{0.5}$ thin films deposited at different $T_S$ originate from the variation in the crystal structure and especially the tetrahedral coordination of iron. In addition to the values of $a$ and $c$, the determination of the all-important Fe-Se tetrahedrons requires determination of the structural parameter, $z$, which can be obtained from the integrated intensities $I_{00l}$ of the (00$l$) diffraction peaks, especially that of the (003) peak. For constant sample illumination area, the x-ray $\theta$-$2\theta$ integrated intensity $I_{00l}$ for a thin film can be expressed as follows: [23]

$$I_{00l} = A \cdot f(p) R_{\parallel}(\theta)(1+\cos^2 2\theta)\left|F(z)_{00l}\right|^2 (1-\exp(-\frac{2\mu t}{\sin \theta})) \qquad (1)$$

where $A$ is a pre-factor which is independent of $l$ and $\theta$, $f(p)$ is the dynamic correction factor [23], $R_{\parallel}(\theta)$ is a sensitivity factor for $\theta$-$2\theta$ scans [24], $F(z)_{00l}$ is the structure factor, $\mu$ is the linear attenuation coefficient, and $t$ is the thickness of the thin films. We also carried out $E_i$=30.5 meV neutron diffraction on two of the samples. Neutrons are penetrating and here not subject to dynamic corrections and absorption correction so the corresponding expression for rocking scan integrated intensity reads: $I_{00l} = A \cdot R_{\perp}(\theta)\left|F(z)_{00l}\right|^2$. To extract $z$ from the neutron and x-ray diffraction data we minimized the profile parameter $R_{wp}$ by varying $A$ and $z$:

$$R_{wp} = [\frac{\sum_{l=1}^{4}(I_{00l}^{obs.} - I_{00l}^{cal.}(z))^2}{\sum_{l=1}^{4}(I_{00l}^{obs.})^2}]^{1/2} \qquad (2)$$

The top right inset to Fig. 3(a) shows observed versus fitted integrated intensities for neutron and x-ray diffraction from the $T_s$=350 °C sample. Clustered around the diagonal and with $R_{wp}$ values of 6% and 17% respectively, the data are consistent with the diffraction models employed. The values of $z$, however, are not consistent to within error bars: $z$(350 °C, neutron)=0.263(2) whereas $z$(350 °C, x-ray)=0.283(3). This indicates that the relatively small number of peaks that could be detected from the thin film samples is insufficient to obtain reliable absolute $z$-values. However, the bottom left inset to Fig. 3(a) shows that the relative integrated x-ray intensity $I_{003}/I_{001}$ depends sensitively on $\Delta z$, the change in $z$ with respect to a chosen reference sample. Assuming that all relevant sample dependence of the integrated intensity is associated with $\Delta z$, we extracted this quantity by fitting such data in an analysis that now is completely independent of the pre-factors in Eq. (1). The consistency of neutron and x-ray data in this analysis provides confidence that the values extracted for $\Delta z$ are reliable.

We now discuss $c$, $a$, and $\Delta z$ which characterize strain induced changes at the atomic scale. While $c$ increases with decreasing substrate temperature (Fig. 3a) the height $\Delta h_{Se/Te}=c\cdot\Delta z+z\cdot\Delta c$ decreases monotonically (Fig. 3b) by a total of 0.08(2) Å through the range of substrate temperatures examined. For $c > 6.02$ Å (with $T_S < 280$ °C), the $z$-value continues to decrease though overlap of the (003) peak with the much more intense MgO(200) peak (Fig. 1a) eventually reduces the experimental accuracy. Thus, in response to a compressive biaxial stress in the basal plane via epitaxy, while the lattice constant $c$ expands, the height $2zc$ of the tetrahedron in the $c$-direction actually *contracts*. This is opposite to the behavior under hydrostatic pressure [18]. As a consequence, the interlayer distance $d_{Se-Se}$ increases with $c$ as shown in Fig. 3b. The overall transport properties of the $FeSe_{0.5}Te_{0.5}$ films vary systematically with $c$ and the corresponding changes in tetrahedral Fe-Se/Te coordination. This is quantitatively displayed in Fig. 3c by plotting $RR = R^{onset}/R(T=290K)$ (where $R^{onset}$ is defined as the value at either $T_C^{onset}$, or T=5K) as the open circles, which vary systematically with $c$. For samples with $c < 5.92$ Å (at high $T_S$) and $c > 6.02$ Å (at low $T_S$) the samples show semiconducting behavior at low temperature with $RR > 1$. In between lie the superconducting samples for which the value of $T_C^{onset}$ distinctively anti-correlates with $RR$ showing a maximum at $c \approx 6$ Å. Rather than directly associated with the changes in $c$, this variation of superconductivity may be linked to changes in the iron coordinating Se/Te tetrahedron.

One characteristic of possible importance is the Fe-As(Se)-Fe bond angle $\alpha$, which for an ideal and undistorted tetrahedron is 109.5°. This angle is indeed realized in the optimally doped and highest $T_c$ "1111" and "122" Fe-SCs [10]. However, in bulk FeSe with $T_C \approx 8$ K, $\alpha$ is less at 104°, and in bulk $FeSe_{0.5}Te_{0.5}$ with $T_C \approx 12$ K $\alpha$ is even smaller at about 99° [10]. In the high-pressure measurements of bulk FeSe [18], $\alpha$ is only slightly reduced, yet $T_C$ is greatly enhanced to 37 K. In the present case of epitaxial thin films of a single composition $FeSe_{0.5}Te_{0.5}$, because both $a$ and $zc$ decrease with increasing $c$, the bond angle $\alpha$ remains roughly unchanged at about 98°, while the superconductivity varies greatly. These results suggest that the bond angle $\alpha$ does not significantly influence $T_C$.

The other two structural characteristics of importance are the chalcogen height $h_{Se/Te} = zc$ of the tetrahedron and Se-Se separation $d_{Se-Se}$. For increasing $c$, the values of $z$ and $h_{Se-Te}$ decrease monotonically. As a result, both the values of $d_{Se-Se}$ (Fig. 3b, open circles) and $c_{Se-Se}$ (not shown) increase monotonically with $c$. Hence under compressive biaxial stress in the basal plane, while the *c*-axis expands, the tetrahedron undergoes compressive distortion such that the essential FeSe(Te) plane becomes "thinner" and further separated from its neighboring planes. Concomitant to the structural variations is the evolution of superconductivity. As shown in Fig. 3d, with decreasing $\Delta h_{Se/Te}$, $T_C^{onset}$ first appears at $\Delta h_{Se/Te} \approx 0.05$ Å, a critical value for superconductivity. As $\Delta h_{Se/Te}$ further decreases, the value of $T_C^{onset}$ increases and reaches a maximum at $\Delta h_{Se/Te} \sim 0$ Å - the reference sample which we expect to be similar to bulk $FeSe_{0.493}Te_{0.507}$ where $z=0.2719(5)$ [25]. This is consistent with the theoretical prediction [9] that as $h_{Se/Te}$ decreases, $q=(\pi+\delta,0)$ spin fluctuations are suppressed in favor of $(\pi,\pi)$ spin fluctuations which promote superconductivity[26-28]. Doping experiments on $FeSe_{1-x}Te_x$ [10] and high-pressure measurements on FeSe [18] also indicate that reducing $h_{Se/Te}$ enhances $T_C$.

In high-pressure experiment on FeSe, all structural parameters decrease with increasing pressure. A smaller chalcogen height $h_{Se}$ accompanies a smaller *c* axis and smaller $d_{Se-Se}$. In contrast, a smaller $h_{Se/Te}$ accompanies a larger *c* axis and a larger interlayer distance in thin films. With decreasing $h_{Se/Te}$ and increasing $d_{Se-Se}$ the structure becomes increasingly two-dimensional and this degrades superconductivity. This is again consistent with our result that $T_C$ exhibits a global maximum versus $\Delta h_{Se/Te}$ decreasing as the structure becomes more two dimensional.

We have observed a systematic variation of $T_C$ with the structural characteristics of the tetrahedron in $FeSe_{0.5}Te_{0.5}$ thin films, though with no $T_C$ enhancement compared to ambient conditions. In contrast, in high-pressure measurements, FeSe, was spectacularly enhanced from 8 K to 37 K [18, 19]. One of the key differences is shown in Fig. 1c, 1d and 3b, where the variations of *a* and $h_{Se/Te}$ with *c* are exactly opposite in thin films and high-pressure measurements [18].

In summary, we show that the substrate temperature can be exploited to apply well defined stress to epitaxial FeSe$_{0.5}$Te$_{0.5}$ thin films on (100)MgO, where the lattice parameters, and more importantly, the Fe-Se(Te) tetrahedron are systematically varied within a single composition. Our results indicate that the superconducting transition temperatures of FeSe$_{0.5}$Te$_{0.5}$ thin films are directly linked to the chalcogen height. This height is in turn essential for ($\pi,\pi$) spin fluctuations, which promote superconductivity, and the Se-Se interlayer distance, which affects the dimensionality of the layered materials.

The assistance of William Ratcliff at the NCNR is gratefully acknowledged. Work by SXH and CLC is supported by the NSF under DMR-0520491 and work by VT and CB is supported by the DOE under DE-FG02-08ER46544.

Fig.1. (Color online) (a) XRD patterns [normalized to (001) peak] of FeSe$_{0.5}$Te$_{0.5}$ thin films deposited at 350 °C (blue line) and 500 °C (red line) on (100)MgO substrates. Inset: tetragonal structure of FeSe$_{1-x}$Te$_x$ with lattice constants $a$ and $c$, Se/Te position $cz$ and bond angle $\alpha$, (b) Lattice constant $c$ (solid symbols) and $a$ (open symbols) of FeSe$_{0.5}$Te$_{0.5}$ thin films as a function of substrate temperature $T_S$. Square symbols are from neutron diffraction. (c) Variation of lattice constant $a$ and $c$ of FeSe under high pressure (ref. [18]) (d) Variation of lattice constant $a$ and $c$ of FeSe$_{0.5}$Te$_{0.5}$ thin films (solid circles, this work), FeSe$_{0.5}$Te$_{0.5}$ target (open circle, this work) and bulk FeSe$_{1-x}$Te$_x$ (solid squares, ref. [21], open squares, ref. [22]). (e) Normalized resistance as a function of temperature for FeSe$_{0.5}$Te$_{0.5}$ thin films deposited at various temperatures $T_S$. Inset: Resistance (normalized to that at T=20K) for thin films at various $T_S$ as indicated.

Fig.2. (Color online) In-plane φ scan (left) and corresponding pole figures (right) about the (101) peak of FeSe$_{0.5}$Te$_{0.5}$ thin films deposited at various substrate temperatures.

Fig.3. (Color online) Variation of structural parameters (a) $\Delta z$, (b) Chalcogen height $\Delta h_{Se/Te}$ (Solid circles) and Se-Se distance $\Delta d_{Se-Se}$ (Open circles) for samples synthesized with different substrate temperatures $T_s$ relative to the reference sample $T_r$=310 °C: $\Delta z(T_S)=z(T_S)-z(T_r)$ etc. Square symbol is from neutron diffraction ($z$(500 °C,neutron)-$z$(350 °C,neutron)+$\Delta z$(350 °C,x-ray)). (c) Superconducting onset transition temperature $T_C^{onset}$ (solid circles) and resistance ratio $RR$ (open circles, see text for definition) of FeSe$_{0.5}$Te$_{0.5}$ thin films as a function of lattice constant $c$. Top right inset in (a) shows the measured versus modeled neutron and x-ray integrated intensity for the $T_s$=350 °C sample. Bottom left inset shows the measured (square) and modeled (line) relative integrated x-ray diffraction intensity $I_{003}/I_{001}$ as a function of $\Delta z$. (d) Superconducting onset transition temperature $T_C^{onset}$ as a function of chalcogen height $\Delta h_{Se/Te}$ relative to the reference sample. All dashed lines are guides to eyes.

**Fig. 1.**

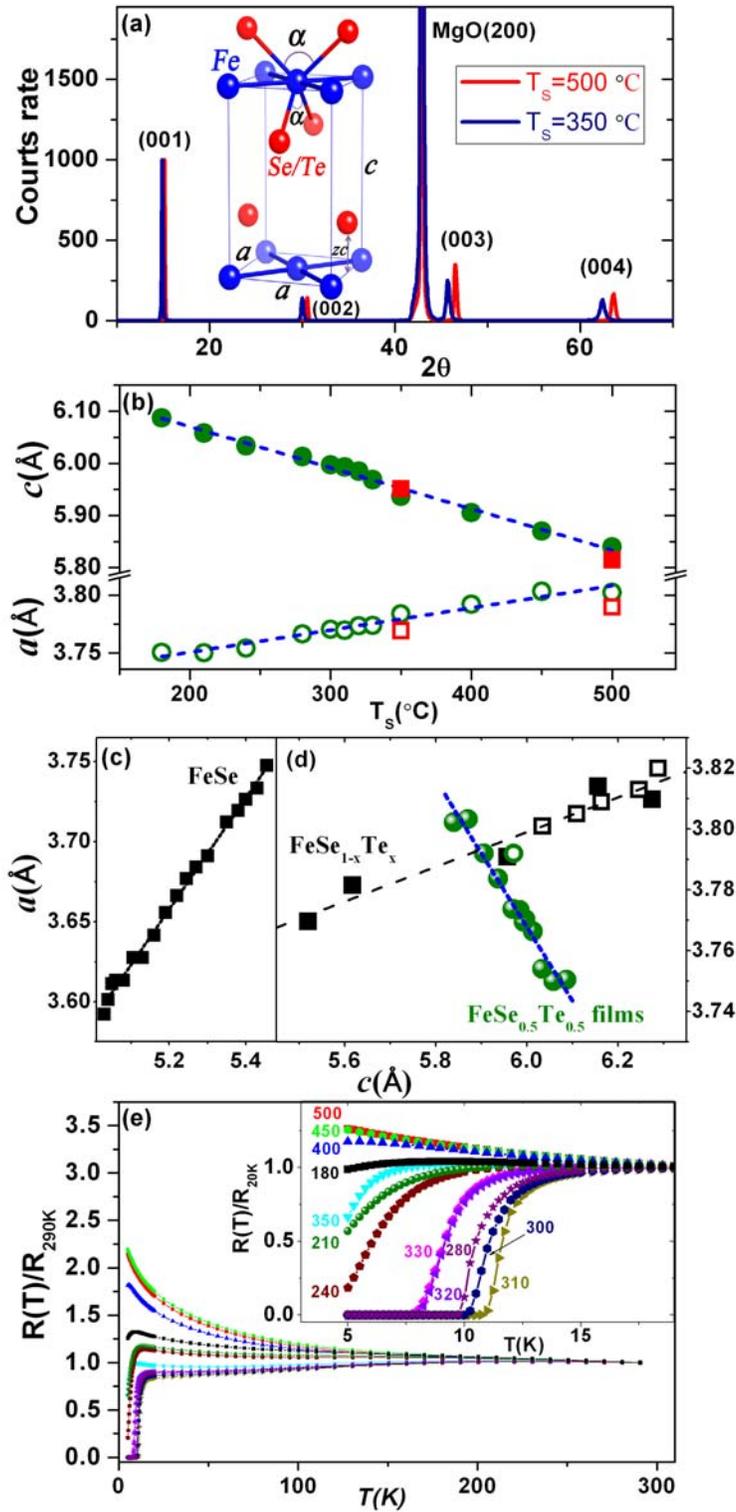

**Fig. 2.**

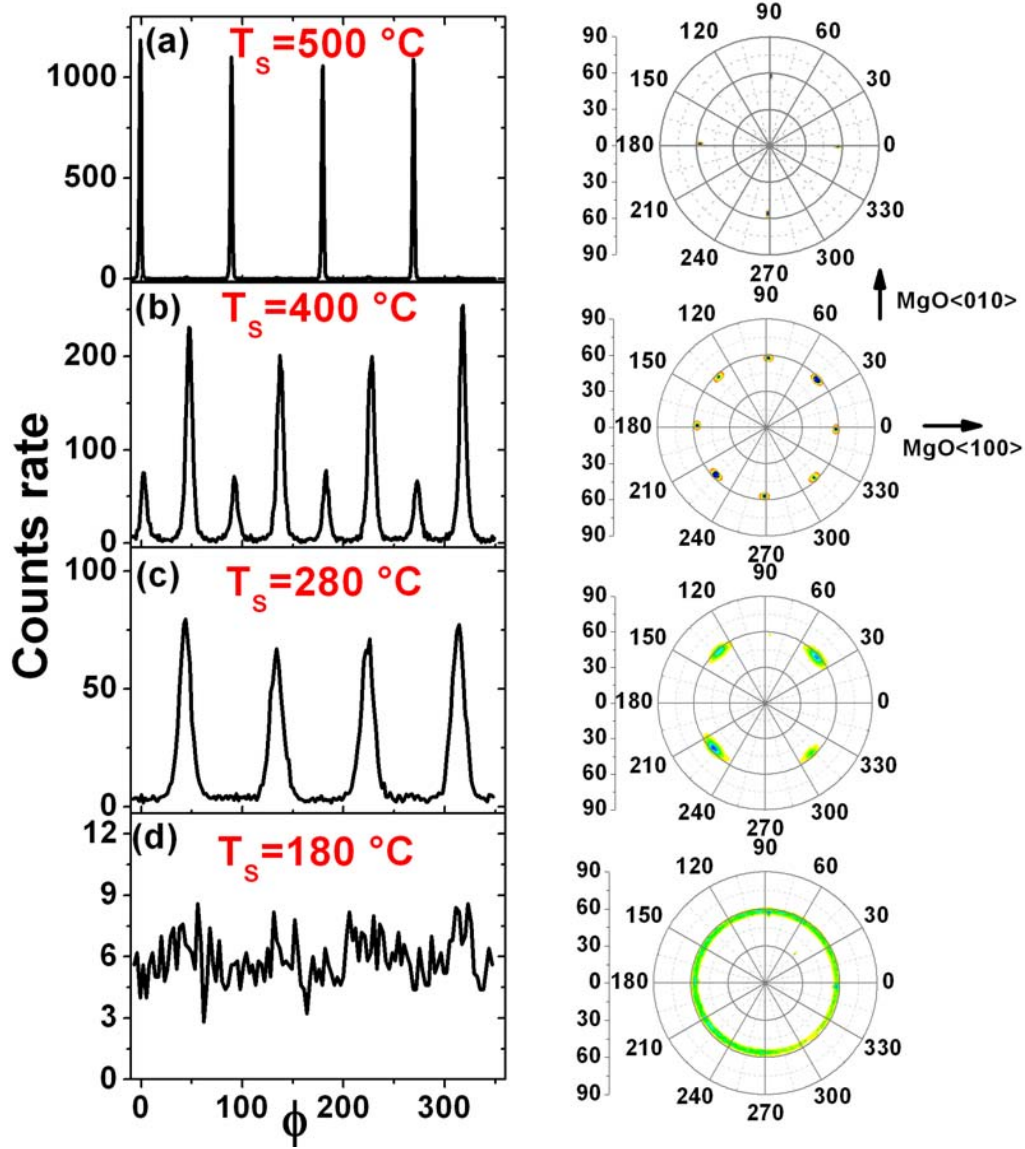

Fig. 3.

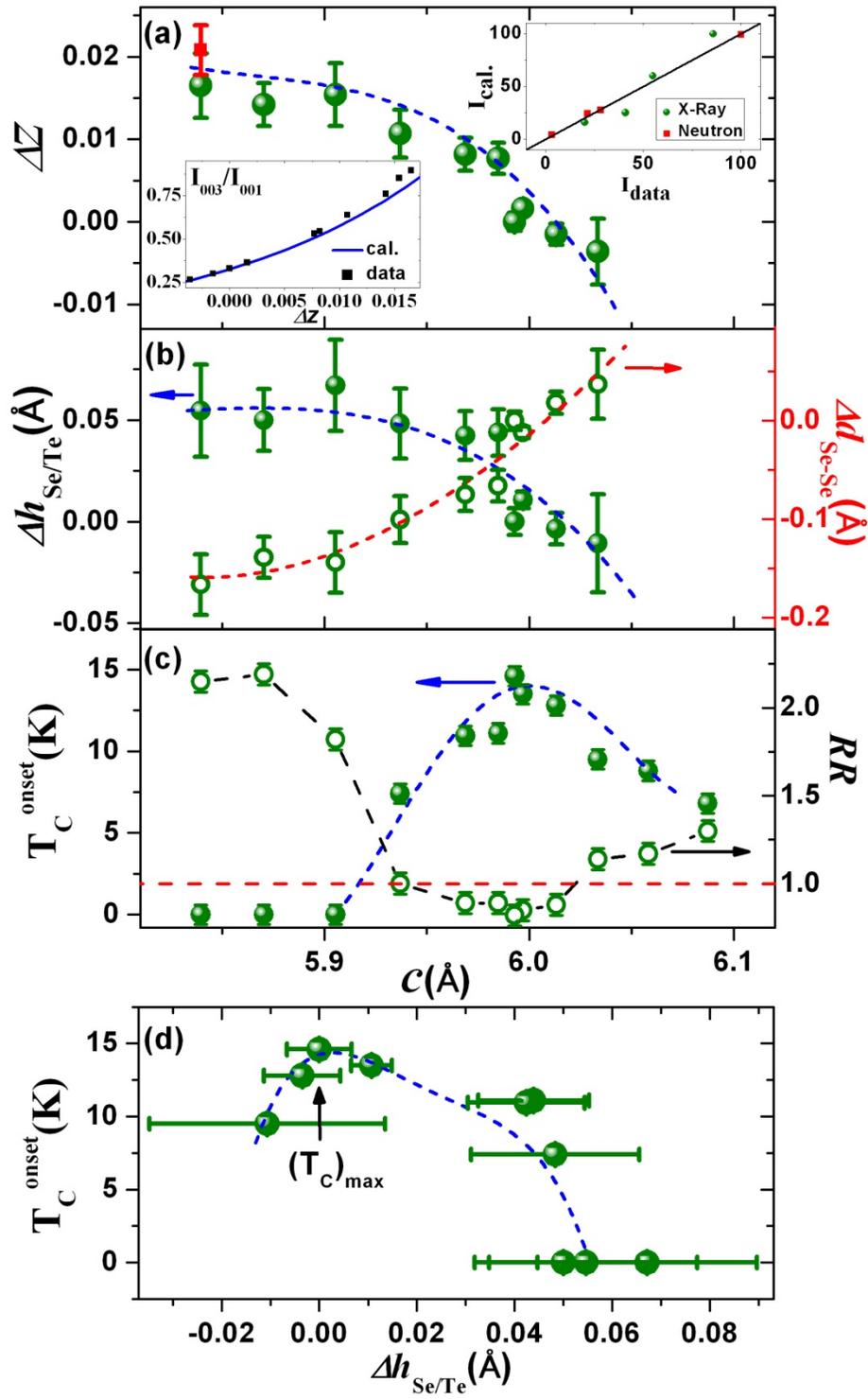